\documentclass[aps,prb,superscriptaddress]{revtex4}
\usepackage{amsmath}
\usepackage{amsfonts}
\usepackage{amssymb}
\usepackage{epsfig}
\usepackage{graphicx}

\begin{document}

\title{Characterizing dynamic length scales in glass-forming liquids}

\author{Walter Kob}
\affiliation{Laboratoire Charles Coulomb,
UMR CNRS 5221, Universit{\'e} Montpellier 2,
34095 Montpellier, France}

\author{S\'andalo Rold\'an-Vargas}
\affiliation{Dipartimento di Fisica, Universit\`{a} di Roma La Sapienza, Piazzale Aldo
Moro 5, 00185 Roma, Italy
}
\affiliation{Departamento de F\'{i}sica
Aplicada, Grupo de F\'{i}sica de Fluidos y Biocoloides, Universidad de
Granada, 18071 Granada, Spain}

\author{Ludovic Berthier}
\affiliation{Laboratoire Charles Coulomb,
UMR CNRS 5221, Universit{\'e} Montpellier 2,
34095 Montpellier, France}

\date{\today}

\maketitle

In their correspondence [1] Flenner and Szamel (FS) compare the temperature
dependence of an alternative dynamic length scale, $\xi_4$, with the
one of $\xi^{dyn}$ that we have studied in [2]. Using computer simulations
of the same system, they conclude that these two  length scales have
a different temperature dependence. In particular, the former does not
follow the striking nonmonotonic temperature dependence we reported for
the latter.  While both types of measurements aim at quantifying the
spatial extent of dynamic correlations in supercooled liquids, the two
procedures differ on essential points, which we now discuss.

While we measured up to what distance the value of the relaxation
time is affected by the presence of an amorphous wall [2], FS quantify
instead the spatial extent of spontaneous dynamic fluctuations at low
wave-vectors for a {\it fixed} timescale (the bulk relaxation
time). These two measurements need not be directly related, although they
seem to coincide at moderate temperatures (see Fig. 1a of FS). While
four-point functions as measured by FS have played a pivotal role in
previous analysis of dynamic heterogeneity, theoretical work has also
revealed a number of shortcomings. Most notably, four-point functions
display a strong dependence on the statistical ensemble chosen to
perform measurements. As a result, they receive distinct contributions
from density and energy fluctuations and show, even in idealized cases,
complex scaling properties [3] which complicates the direct extraction of
a correlation length. A second difficulty lies in the fact that these
various contributions have different temperature dependences, with
a crossover taking place very close to the mode-coupling temperature
where the subtle effects reported in [3] occur. We remark that this
corresponds to the temperature scale where $\xi_4$ and $\xi^{dyn}$ start
to differ. These known weaknesses of four-point functions had in fact
motivated our study of an alternative correlation length scale that is
free of such ambiguities.

Given these important differences, it is not clear how a nonmonotonic
temperature evolution of dynamic correlations will manifest itself
in the numerical data of FS. While FS argue that the evolution of
$\xi_4$ with the relaxation time presents a crossover (Fig.1c in FS),
we point out that the shown data does not clearly show such a change
in behavior. Another possibility, not explored by FS, could be that the
functional form of the correlator $S_4(q,t)$ changes with temperature,
in agreement with the idea that the geometry of the relaxing entities
changes with temperature, as argued in [2]. Since the reported effect
is small (Fig. 2b in [2]), one would presumably need a relative accuracy
of $S_4(q,t)$ of better than 1\% at low wave-vectors. The data shown by
FS demonstrates that at present this remains technically extremely difficult.

Finally, the crossover we report also coincides with the emergence of
nontrivial static correlations in the system, a result that cannot be
obtained using purely dynamic correlations. Overall, this suggests that
the approach of pinning particles, for which no a priori knowledge of
the nature of the relaxing entities is needed, is extremely helpful
to study small but important effects in the relaxation properties of
glass-forming systems that cannot be drawn from the measurements of
FS. We point out that since the publication of [2], we independently
confirmed the qualitative change of transport properties in our system
near the mode-coupling crossover by analyzing the temperature evolution
of dynamical finite size effects [4].

\vspace*{10mm}

References:

[1] E. Flenner and G. Szamel;
Characterizing dynamic length scales in glass-forming liquids,
Comment to appear in Nature Physics, arXiv:12xx.xxxx

[2] W. Kob, S. Roldan-Vargas, and L. Berthier;
Non-monotonic temperature evolution of dynamic correlations in
glass-forming liquids,
Nature Phys. {\bf 8}, 164 (2012)

[3] L. Berthier, G. Biroli, J.-P. Bouchaud, W. Kob, K. Miyazaki, D. Reichman;
Spontaneous and induced dynamic fluctuations in glass-formers I:
General results and dependence on ensemble and dynamics,
J. Chem. Phys. {\bf 126}, 184503 (2007)

[4] L. Berthier, G. Biroli, D. Coslovich, W. Kob, and C. Toninelli;
Finite size effects in the dynamics of glass-forming liquids,
arXiv:1203.3392, to appear in Phys. Rev. E (2012) 

\end{document}